\documentclass[%
pra,
twocolumn,
superscriptaddress,
amsmath,amssymb,groupaddress,longbibliography,
aps,
]{revtex4-2}
\usepackage[colorlinks=true,linkcolor=blue,urlcolor=blue,citecolor=blue,pdfusetitle]{hyperref}

\input{Qcircuit}
\usepackage{physics,palatino}
\usepackage{graphicx}% Include figure files
\usepackage{dcolumn}% Align table columns on decimal point
\usepackage{bm}% bold math
\newcommand{\Id}{\mathbb{I}}
\newcommand{\ui}{\mathrm{i}}
\usepackage[normalem]{ulem}
\usepackage[dvipsnames]{xcolor}

\usepackage{tikz}
\usetikzlibrary{decorations.markings,arrows,arrows.meta}

\makeatletter
\newcommand{\thickhline}{%
    \noalign {\ifnum 0=`}\fi \hrule height 1pt
    \futurelet \reserved@a \@xhline
}
\newcolumntype{"}{@{\hskip\tabcolsep\vrule width 1pt\hskip\tabcolsep}}
\makeatother

\usepackage{pifont}
\newcommand{\cmark}{\ding{51}}%
\newcommand{\xmark}{\ding{55}}%

\begin{document}
 
%\preprint{APS/123-QED}
 
\title{Distributing quantum correlations through local operations and classical resources}% Force line breaks with \\
 
\author{Adam G. Hawkins}
\affiliation{Centre for Quantum Materials and Technologies, School of Mathematics and Physics, Queen’s University Belfast, BT7 1NN Belfast, UK}

\author{Hannah McAleese}
\affiliation{Centre for Quantum Materials and Technologies, School of Mathematics and Physics, Queen’s University Belfast, BT7 1NN Belfast, UK}
 
\author{Mauro Paternostro}
\affiliation{Quantum Theory Group, Dipartimento di Fisica e Chimica - Emilio Segr\`e, Universit\`a degli Studi di Palermo, via Archirafi 36, I-90123 Palermo, Italy}
\affiliation{Centre for Quantum Materials and Technologies, School of Mathematics and Physics, Queen’s University Belfast, BT7 1NN Belfast, UK}

\date{\today}% It is always \today, today,
 
\begin{abstract}
Distributing quantum correlations to each node of a network is a key aspect of quantum networking. Here, we present a robust, physically motivated protocol by which global quantum correlations, as characterized by the discord, can be distributed to quantum memories using a mixed state of information carriers which possesses only classical correlations. In addition, such distribution is done using only bilocal unitary operations and projective measurements, with the degree of discord being measurement-outcome independent. We explore the scaling of the performance of the proposed protocol with the size of the network and illustrate the structure of quantum correlations that are shared by the nodes, showing its dependence on the local operations performed. Finally, we find the counterintuitive result that even more discord can be generated when the resource state undergoes correlated dephasing noise, allowing high fidelities with mixtures of the Bell basis such as Werner states.
\end{abstract}

\maketitle
 
\section{Introduction}

In recent times, the idea of a quantum internet \cite{KimbleReview,WehnerReview} has become one of the most sought-after uses of quantum resources in modern quantum information theory.
In particular, the leveraging of quantum entanglement as a resource to perform communication in a way that is not allowed in classical communication is paramount to gaining quantum advantage.
Entanglement is often referred to by the blanket term ``quantum correlations", however, with the concept of quantum discord \cite{Ollivier,Henderson}, one can only consider entanglement to be a subset of what can be labelled as quantum (or ``non-classical") correlations.
In fact, it is possible for states to possess non-zero quantum correlations while simultaneously exhibiting zero entanglement \cite{Qasimi}, and so measures of entanglement do not, in general, capture all of the quantum correlations of a given system.

The full usefulness of said states with regards to quantum information processing remains to be seen.
However, it has been shown that a state with non-zero discord can always be used to `activate' bipartite entanglement between a system and a corresponding auxiliary; hence, discord is a resource in quantum information processing \cite{Piani}. For certain separable non-classical states, it has been shown that this entanglement can be localized to the original system via stochastic entanglement swapping \cite{GharibianQuantumness}.
Furthermore, it has been established that discord is a resource for a number of applications, such as quantum computation \cite{Datta,Lanyon_QC}, remote state preparation \cite{Dakic}, state discrimination \cite{Roa_2011,Li_2012,Pang_2013,Zhang_2013} and quantum cryptography \cite{Pirandola_discord,Wang}.

Just as with entanglement, if quantum discord is to be used as a resource in quantum networking, then methods for distributing it to quantum memories at each node of the network must be established.
For entanglement to be distributed, either the entanglement can be generated initially to then be sent directly to the memories, or some sort of non-local operation(s) must be introduced to the memories.
However, it has been shown that it is possible to generate discord via \textit{local}, non-unitary channels acting on one or more of the subsystems in a state \cite{StreltsovIncrease,CiccarelloChannels,Gokhan}, with multiple physical examples of said discord generation being detailed \cite{CampbellProp,Ciccarello,Lanyon}. These discordant states often have asymmetric structures, depending on the channel which is applied.
If discord is distributed to quantum memories in a network, the possibility of entanglement distribution to these discordant states via interactions with a \textit{separable} carrier state then arises~\cite{Cubitt,Kay,McAleese,PhysRevResearch.6.033317}. Thus, discord is a resource in entanglement distribution \cite{Streltsov_Cost,Chuan}.

Here, we take the idea of locally generating discord and outline a robust discord distribution protocol.
This protocol involves a source of information carriers, which are initially in a mixed, classically correlated state, being used to distribute significant amounts of discord to spatially-separated quantum memories. This is done via local unitary interactions (with respect to the memories) and projective measurements only.
We show here that this distribution of discord does not depend on the measurement outcomes of said projective measurements; only the basis used to measure the information carriers determines the amount of distributed discord.
In fact, we detail a modified version of this protocol whereby a discordant resource state of a quantum key distribution (QKD) protocol can be generated with unit probability.
Our protocol not only appears to scale well with the number of memories in the network, but also exhibits robustness to the preparation and measurement apparatus used.
Moreover, we highlight how the protocol can perform even better under certain correlated noise models.

In Sec.~\ref{sec: protocol}, we describe the bipartite protocol and describe the quantum channel acting on the memories as a result of the protocol, while also outlining different types of discordant states.
% We analyse this channel in order to determine why it is able to distribute discord to memories which are initially uncorrelated, with reference to unital and semiclassical channels.
We then investigate the robustness of the bipartite protocol in Sec.~\ref{sec: robustness} with respect to measurement apparatus as well as the initial carrier and memory states. We further evaluate how our chosen quantifier of the quantum correlations affects how much discord is present in the final state, while also seeking numerical evidence of the true resource of the protocol. In Sec.~\ref{sec: system sizes}, we consider protocols involving more memories, and calculate how the amount of distributed discord scales with the number of nodes in the network.
We also offer a benchmark to gauge how significant the amount of distributed discord is for each system size. In Sec.~\ref{sec: discord structure}, we explore the structure of the discordant final states of the memories upon the conclusion of the protocol, and how such structure depends on the local operations performed at each node. We consider and illustrate this for multiple system sizes in order to establish structural patterns and consistencies. We present our conclusions in Sec.~\ref{sec: conclusions} and discuss potential future implementations of our protocol.
 
\section{The bipartite protocol}\label{sec: protocol}
Consider a source which outputs a pair of {\it carrier} qubits (labelled as $C_1$ and $C_2$ respectively) prepared in the mixed state
\begin{equation}\label{eq: initial carriers}
    \rho_{C_1C_2} = \frac{1}{2}(\ketbra{++}{++}+\ketbra{--}{--})_{C_1C_2}\,,
\end{equation}
where $\ket{\pm} = (\ket{0}\pm\ket{1})/\sqrt{2}$ is the eigenstate of the $x$ Pauli operator with eigenvalue $+1$, and $\{\ket{0},\ket{1}\}$ is the computational basis of eigenstates of the $z$ Pauli operator $Z$ such that $Z\ket{0}=\ket{0}$ and $Z\ket{1}=-\ket{1}$. Eq.~\eqref{eq: initial carriers} obviously carries no quantum correlations. Each carrier is sent to a respective quantum memory, which we assume to be spatially separated by an arbitrary distance. The memories $M_1$ and $M_2$ are initially in the {pure} product state $\rho_{M_1M_2} = \ketbra{++}{++}_{M_1M_2}$.

A controlled-$Z$ gate $U_{Z,i}=\ketbra{0}{0}_{C_i}\otimes\openone_{M_i}+\ketbra{1}{1}_{C_i}\otimes Z_{M_i}$ is then performed over each carrier-memory pair $\{C_i,M_i\}~(i=1,2)${, where $\openone$ is the identity}. The carriers are then locally measured in a suitable orthonormal basis $\{\ket{\psi}_{C_i},\ket{\psi^\perp}_{C_i}\}$ with %in the hopes that quantum correlations can be established between the memories.
% \begin{figure}
%     \centering
%     \includegraphics[width=\linewidth]{Discord Dis Paper FIG.pdf}
%     \caption{Diagram displaying each step of the protocol. (1) A photon source emits photons in a mixed state, such that they have classical correlations but no quantum correlations, and are each sent to a spatially-separated atom suspended in an optical cavity. (2) Each photon is reflected off their respective cavity, realising a \textsc{cphase} operation. (3) The photons are then each locally measured in some polarisation basis. There is now, in general, resulting discord between the atoms.}
%     \label{fig: protocol diagram}
% \end{figure}
%We model the measurements made on the carriers by the generalised projectors
\begin{equation}
%    \ketbra{\psi_i}{\psi_i}\mathrm{\ with\ }
\ket{\psi}_{C_i}=\cos{\frac{\theta_i}{2}}\ket{0}_{C_i} + e^{\ui\phi_i}\sin{\frac{\theta_i}{2}}\ket{1}_{C_i}
\end{equation}
and ${}_{C_i}\!\braket{\psi^\perp}{\psi}_{C_i}=0$. We have introduced the Bloch angles $\theta_i \in[0,\pi]$ and $\phi_i \in[0,2\pi)$ for the carrier $C_i$. Such simple protocol can be presented diagrammatically by the following quantum circuit %in Fig.~\ref{fig: circuit diagram}.
$$
\Qcircuit @C=3.7em @R=.15em @! {
&\ctrl{4}& \qw&\meter \\
\lstick{\rho_{C_1C_2}\,\,} &&&\\
&\qw & \ctrl{4}&\meter\\
&&&\\
 \lstick{\ket{+}_{M_1}}&\gate{Z} &\qw&\qw\\
 &&&\rstick{\,\,\rho'_{M_1M_2}}\\ 
 \lstick{\ket{+}_{M_2}}&\qw& \gate{Z}&\qw
 \gategroup{1}{1}{3}{2}{.5em}{\{}
  \gategroup{5}{3}{7}{4}{.7em}{\}}
}
$$
%\begin{figure}[ht]
  %  \centering
    %\includegraphics[width=0.8\linewidth]{circuit_diagram.pdf}
   % \caption{Quantum circuit illustrating the discord-distribution protocol.}
    %\label{fig: circuit diagram}
%\end{figure}
where we have used the standard symbols for the $U_{Z}$ gate. We are interested in the reduced state of the memories upon application of the protocol
\begin{equation}
\rho'_{M_1M_2}=\mathcal{N}\Tr_{C_1C_2}\left[\Pi_\psi U_{CM}\left(\rho_{C_1C_2}\otimes\rho_{M_1M_2}\right)U_{CM}^\dag\Pi_\psi\right]
\end{equation}
% \begin{equation}
%     \rho'_{M_1M_2}=\Tr_{C_1C_2}\left[U_{CM}\left(\rho_{C_1C_2}\otimes\rho_{M_1M_2}\right)U_{CM}^\dag\right]
% \end{equation}
{with $U_{CM}=U_{Z,1}\otimes U_{Z,2}$, $\Pi_\psi=\ketbra{\psi}{\psi}_{C_1}\otimes\ketbra{\psi}{\psi}_{C_2}$ and $\mathcal{N}$ denoting a normalization constant.}% scalar found by taking the reciprocal of the trace of the post-measurement state.}
We stress that we cannot interpret the evolved state of the memories as the output of individual quantum channels applied to $M_1$ and $M_2$ as the carriers are initially classically correlated. We do, however, highlight that the operations performed are \textit{local} with respect to the $M_1:M_2$ partition, i.e., the protocol is \textit{bilocal}. The final reduced state of the memories takes the following form:
%This is apparent from the following form taken by the reduced memory  state
%By considering this protocol to be a channel that acts on the memories, we can consider the Kraus operators which act on these memories. After careful analytical work, one can find that the final, measurement-dependent state of the memories $\rho^\prime_M$ is given by 
%\begin{widetext}
    \begin{equation}\label{eq: final memory state}
        \begin{aligned}
 &   \rho_{M_1M_2}^\prime = \bigotimes_{i=1,2}\left(\cos^2\frac{\theta_i}{2}\ketbra{+}{+}+\sin^2\frac{\theta_i}{2}\ketbra{-}{-}\right)_{M_i}\\
   % &\rho_{M}^\prime = \cos^2\frac{\theta_1}{2} \cos^2\frac{\theta_2}{2} \ketbra{++}{++} + \sin^2\frac{\theta_1}{2}\sin^2\frac{\theta_2}{2}\ketbra{--}{--}
   % + \cos^2\frac{\theta_1}{2} \sin^2\frac{\theta_2}{2}
    %\ketbra{+-}{+-}+ \sin^2\frac{\theta_1}{2} \cos^2\frac{\theta_2}{2}\ketbra{-+}{-+}\\
   &+
   %\frac{\sin{\theta_1}\!\sin{\theta_2}}{4}\!
   \alpha(\theta_1,\theta_2)
   \left(e^{\ui\phi_+}\ketbra{++}{--} {+} {e^{\ui\phi_-}}\ketbra{+-}{-+} {+} \mathrm{h.c.}\right)_{M_1M_2}
        \end{aligned}
    \end{equation}
%\end{widetext}
with $\alpha(\theta_1,\theta_2) = (\sin\theta_1 \sin\theta_2) /4$ and $\phi_\pm=\phi_1\pm\phi_2$. For suitable choices of the parameters of the projections performed on the carriers, Eq.~\eqref{eq: final memory state} might bring about quantum correlations. We include a brief analysis of this protocol -- with respect to the quantum channel involved -- as an appendix to this paper.

\begin{figure}
    \centering
    \includegraphics[width=0.7\linewidth]{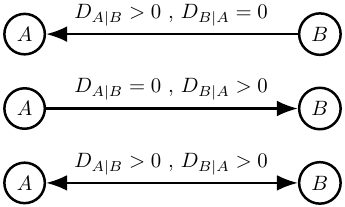}
    \caption{Graphical representations of possible discord structures between two arbitrary systems $A$ and $B$. In the bottom diagram, $D_{A\vert B}$ is not necessarily equal to $D_{B\vert A}$. We present {\it classical-quantum}, {\it quantum-classical}, and {\it quantum-quantum} states in a top-to-bottom configuration.}
    \label{fig: discord structure}
\end{figure}

To quantify the quantum correlations, we choose the quantum discord in its original definition~\cite{Ollivier,Henderson}:
%
%In order to emphasise the message of this section, we first look at the definition of asymmetric bipartite quantum discord as defined in Refs.~\cite{Ollivier,Henderson}. 
Consider a bipartite quantum system prepared in state $\rho_{AB}$. We introduce % the discord is defined as the minimum difference between 
the mutual information $I(A:B)$ -- a measure of the total correlations within a given quantum state -- and total classical correlations $J(A:B)$
\begin{align}
    I(A:B) &= S(\rho_A) + S(\rho_B) - S(\rho_{AB})\,,\\
    J(A:B) &= S(\rho_{A}) - S(\rho_{A\vert B})\,,
\end{align}
where $S(\mu)$ is the von Neumann entropy of the generic state $\mu$. Here, $\rho_{A(B)}=\Tr_{B(A)}[\rho_{AB}]$ and $\rho_{A\vert B}$ are the reduced state of subsystem $A$ $(B)$ and the conditional state of $A$ given a measurement performed on $B$, respectively. The measurement on $B$ is modelled as a set of rank-1 projectors $\{\Pi_j^B\}$ \cite{GalveAlmost} labelled by  the index $j$ of the corresponding measurement outcomes, each occurring  with a probability $p_j = \Tr[\Pi_j^B \rho_{AB}]$. The difference between $I(A:B)$ and $J(A:B)$, minimized over all the projective measurements on $B$, embodies the quantum discord~\cite{Modi}
\begin{equation}
\label{defdiscord}
    D_{A\vert B} = S(\rho_B) - S(\rho_{AB}) + \min_{\{\Pi_j^B\}} \sum_j p_j S(\rho_{AB\vert \Pi_j^B})\,,
\end{equation}
where $\rho_{AB\vert \Pi_j^B}=(\Pi_j^B\rho_{AB}\Pi_j^B)/p_j$ is the state resulting from the application of the measurement operators. A caveat of the above definition is that, in general, $D_{A\vert B}\neq D_{B\vert A}$  due to the asymmetric nature of the quantum conditional entropy \cite{Modi}. In fact, there exist  \textit{classical-quantum} states, particularly those resulting from the action of  local channels, which showcase $D_{A\vert B}>0$ with $D_{B\vert A}=0$~\cite{GharibianQuantumness,CiccarelloChannels}. Examples of such states include the resource state for the QKD protocol dubbed B92~\cite{B92} along with, in general, any state obtained as a result of the protocol proposed here but limited to just one quantum memory~\footnote{We use the term {\it in general} to remark that there exist certain carrier measurement bases, such as the computational one, whereby no discord is generated. The resulting state will still have the same structure as a classical-quantum state, but will not possess asymmetric discord. Therefore, the state, in general, has asymmetric discord. %\HM{Could we rewrite this last sentence?: When using most measurement bases, however, the state has asymmetric discord.} 
This reasoning is also applied to the rest of Sec.~\ref{sec: discord structure} in all instances where we refer to a state having discord {\it in general}.}. %This 'one-way' discord arises from the indistinguishable nature of different pure states in a mixed state \cite{CiccarelloChannels,Groisman}. 
The different one- and two-way discord structures are shown graphically in Fig.~\ref{fig: discord structure}.
% \begin{equation*}
% \begin{split}
% &\begin{tikzpicture}[node distance={50mm}, very thick, decoration={
%     markings,
%     mark=at position 0.5 with {\arrow{>}}}, main/.style = {draw, circle}] 
% \node[main] (1) {$A$};
% \node[main] (2) [right of=1] {$B$};
% \draw[-{Latex[scale=1.2]}] (2) -- node[midway, above right, sloped, pos=0.92] {$D_{A\vert B}>0\text{ , }D_{B\vert A}=0$} (1);
% \end{tikzpicture}\\
% &\begin{tikzpicture}[node distance={50mm}, very thick, decoration={
%     markings,
%     mark=at position 0.5 with {\arrow{>}}}, main/.style = {draw, circle}] 
% \node[main] (1) {$A$};
% \node[main] (2) [right of=1] {$B$};
% \draw[-{Latex[scale=1.2]}] (1) -- node[midway, above right, sloped, pos=0.08] {$D_{A\vert B}=0\text{ , }D_{B\vert A}>0$} (2);
% \end{tikzpicture}\\
% &\begin{tikzpicture}[node distance={50mm}, very thick, decoration={
%     markings,
%     mark=at position 0.5 with {\arrow{>}}}, main/.style = {draw, circle}] 
% \node[main] (1) {$A$};
% \node[main] (2) [right of=1] {$B$};
% \draw[{Latex[scale=1.2]}-{Latex[scale=1.2]}] (1) -- node[midway, above right, sloped, pos=0.08] {$D_{A\vert B}>0\text{ , }D_{B\vert A}>0$} (2);
% \end{tikzpicture}
% \end{split}
% \end{equation*}

When using Eq.~\eqref{defdiscord}, the maximum degree of discord in state \eqref{eq: final memory state} (obtained by optimizing  with respect to the four angles entering it) is found to be $D_{M_1\vert M_2}\simeq0.2018$~\footnote{For reference, the discord of a maximally entangled pure state of two qubits is exactly 1.}, which is also the maximum amount of locally generated discord achievable -- starting with a classically correlated state -- in a bipartite qubit system ~\cite{CiccarelloChannels}. Such result can be achieved by picking up values for the angles $\theta_j$ and $\phi_j$ from a manifold, an interesting example being $\theta_1=\pi/2$, $\theta_2=\pi/4$, for any choice of  $\phi_{1,2}$. While, needless to say, different choices of parameters taken from the optimal manifold would result in different (yet iso-discordant) states, such optimal degree of discord is obtained {\it independently} of the value of the relative phases $\phi_j$ chosen for the measurements on the carriers: All the memory states achieved by projecting the carriers onto the elements of a chosen measurement basis bring about the same degree of discord. %\MP{I do not agree fully with this statement. We still have to postselect the outcome in that we have to project the state of the carriers. To me, postselection-free means that I can trace out the  degrees of freedom of the carriers...the experimentalist should just trash the carriers.....we do not do that. On the contrary, what we are saying is that {\bf measurements over orthogonal states belonging to suitably chosen planes of the Bloch spheres of the carriers lead to the same degree of discord}. In my view, 
In a sense, this implies that the protocol described here actually prepares states belonging to \textit{subspaces} of iso-discordant states. Among such subspaces, there is one whose elements all achieve the degree of discord of the B92 resource state. In practical terms, the data achieved for any detection configuration belonging to such optimal plane could be retained for the sake of distributing maximum discord, and unless a {\it specific}  %Of course, the state obtained will depend on the measurement outcomes, so postselection and/or measurement-dependent local unitary transformations will be needed if a specific 
discordant state is desired. %\MP{do we need this comment? Moreover, there exists only a discrete set of parameters for which the distributed discord is zero; i.e., the discord is, in general, non-zero (see Sec.~\ref{sec: robustness})}.

\subsection{Generating the Resource State for the B92 QKD Protocol}

% \begin{figure*}[ht!]
%     \centering
%     {\bf (a)}\hskip9cm{\bf (b)}\\
%     \includegraphics[width=.5\linewidth]{GQD_meas_params_heatmap.pdf}\quad\includegraphics[width=.47\linewidth]{discord_same_params_mixed.pdf}
%     \caption{{\bf (a)} Distributed bipartite GQD between the memories plotted as a function of the parameters $\theta_{1,2}$ entering the measurements on the carriers. {\bf (b)} A study of the bipartite GQD distributed to the memories against the mixing parameter $\lambda$ in Eq.~\eqref{q: carrier state mixed} and obtained assuming the optimal measurement bases for the carriers for any value of $\lambda$.}
%     \label{fig: heatmap}
% \end{figure*}
\begin{figure*}
    \centering
    \includegraphics[width=\linewidth]{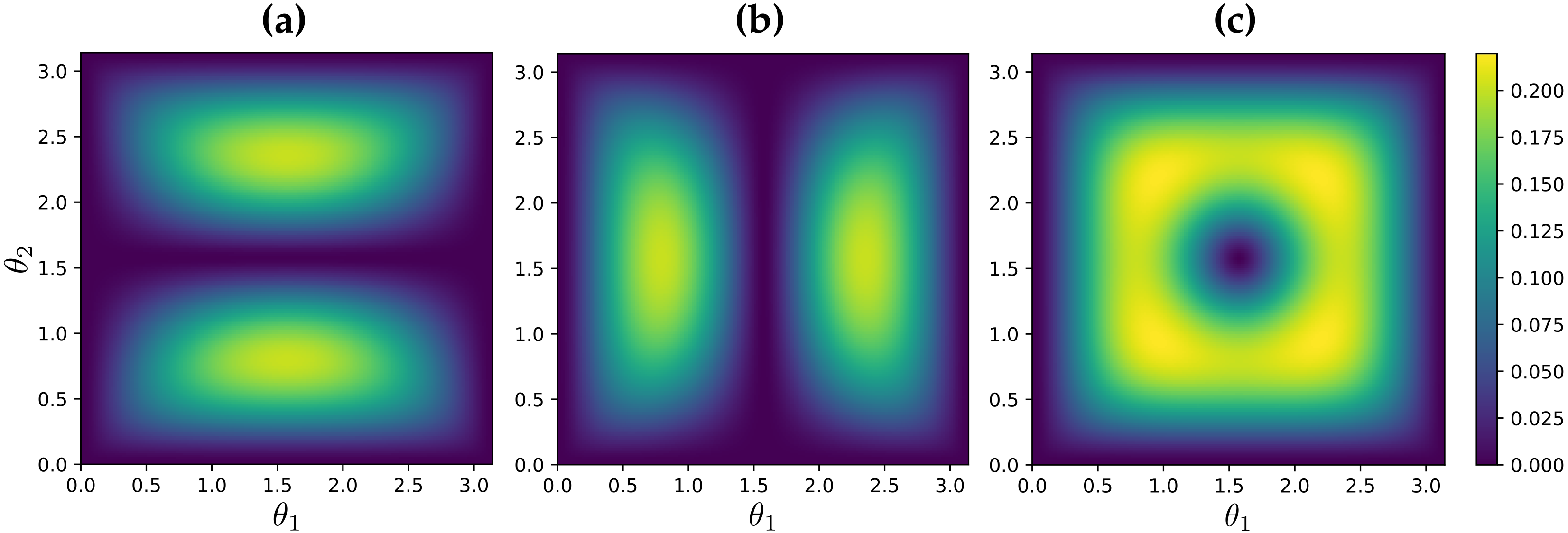}
    \caption{{Distributed quantum correlations in the state $\rho_{M_1M_2}^\prime$ plotted against the parameters $\theta_{1,2}$ entering the measurements on the carriers. Since the amount of quantum correlations is independent of $\phi_{1,2}$, we set $\phi_1=\phi_2=0$ without loss of generality. We consider three non-equivalent quantifiers of discord: panels \textbf{(a)} and \textbf{(b)} show the two asymmetric measures  $D_{M_1\vert M_2}(\rho_{M_1M_2}^\prime)$ and  $D_{M_2\vert M_1}(\rho_{M_1M_2}^\prime)$, respectively, while panel \textbf{(c)} is for the symmetric GQD $\mathcal{G}(\rho_{M_1M_2}^\prime)$.}}
    \label{fig: heatmap}
\end{figure*}

Finally, it is worth noting that this protocol may be rewritten in different bases for the purpose of obtaining different states. For example, by writing the protocol in the computational basis $\{\ket{+},\ket{-}\}\to\{\ket{0},\ket{1}\}$ and using a controlled-Hadamard gate instead of a {controlled-$Z$}, it is possible to achieve the discordant resource state of the B92 QKD protocol \cite{B92} with $100\%$ probability. This is the case when only Bob's carrier {$C_2$} interacts {with $M_2$} ($C_1$ can either just be held by Alice or can be state-mapped onto $M_1$) and is then measured in any basis defined by $\theta_2=\pi/2$ (e.g., $\{\ket{+},\ket{-}\}$), with both measurement outcomes projecting the final state onto the B92 resource state; not only is the amount of discord independent of the measurement outcome but so is the state, which itself is not a function of $\phi_2$. This would mean that, instead of the method outlined in Ref.~\cite{NC}, the protocol can be performed in the case where the resource state is supplied to Alice and Bob by a trusted third party -- without the need for any direct initial classical communication or exchanging of quantum states between themselves. The third party can help perform this protocol while also not knowing which of the pure states in the mixture have been supplied to Alice and Bob. If the third party is untrusted, though, then this method will no longer be secure, as they can know which pure state is supplied with certainty.

One may also consider extending this to an $N$-party conference key agreement (CKA) protocol which, in principle, could generate an identical key for 1 `Alice' and $N-1$ `Bob's (or vice versa). However, any such resource state, albeit discordant, would be completely separable, and it has been shown that any CKA protocol in which a single copy of a multipartite state is distributed to all parties involved requires (biseparable) entanglement to be unconditionally secure \cite{Curty_2004,Carrara_2021}. Furthermore, the B92 protocol performed in this way will have no guarantee of security against adversarial attacks even in the bipartite scheme; if an adversary correctly measures one of the carrier qubits before the controlled-Hadamard interaction in the computational basis then they will know, with certainty, which bit that Alice receives and which quantum state that Bob receives. Therefore, additional methods would need to be considered in order to overcome these vulnerabilities.

\section{Robustness of the Protocol}\label{sec: robustness}
At this point, we introduce a different quantifier of quantum correlations,
namely the \textit{global quantum discord} (GQD) first defined in Ref.~\cite{Rulli} so as to pave the way to and perform an analysis of the generated quantum correlations in protocols involving more than two memories (see Sec.~\ref{sec: system sizes}). {Quantitatively, GQD is defined as 
\begin{equation}
\label{def:gqd}
\begin{split}
{\cal G}(\rho_{\{X\}}) = \min_{\{\Pi_k\}} \Bigg[ S\boldsymbol{(}&\rho_{\{X\}}\,\|\, \Phi(\rho_{\{X\}})\boldsymbol{)}\\
&- \sum_{j=1}^N S\boldsymbol{(}\rho_{X_j}\, \|\, \Phi_j(\rho_{X_j})\boldsymbol{)}\Bigg]\,,
\end{split}
\end{equation}
where $\{X\}=\{X_1,\dots,X_N\}$ stands for the total system comprising each subsystem $X_j$, $S(\mu\,\|\,\tau)$ denotes the quantum relative entropy~\cite{VedralRelative} between two generic states $\mu$ and $\tau$, $\Phi_j(\rho_{\{X\}})= \sum_{j^\prime}\Pi^{X_j}_{j^\prime} \rho_{X_j} \Pi^{X_j}_{j^\prime}$ and $\Phi(\rho_{\{X\}})=\sum_k \Pi_k \rho_{\{X\}}\Pi_k$, with $\Pi_k=\Pi_{j_1}^{X_1}\otimes \cdots \otimes \Pi_{j_N}^{X_N}$ and $k$ denoting the index string $(j_1\cdots j_N)$ \cite{Rulli}. Here, a projector is again represented by the symbol $\Pi$.}
Our choice of figure of merit is due to the fact that our interest when addressing a multipartite memory will move away from bipartitions to address the degree of quantum correlations shared by all the elements of the memory compound. Moreover, GQD  bypasses the intrinsic asymmetry of Eq.~\eqref{defdiscord}{, and we wish to explore how the distributed quantum correlations in the bipartite protocol differ depending on the quantifier.}
%, unlike the definition given in Refs~\cite{Ollivier,Henderson}, this definition is symmetric under the change of subsystem and is defined in terms of relative entropy. Furthermore, since this is a global measure of discord, if there exists no quantum correlations in any of the bipartitions (not tracing any subsystem out) then the GQD is exactly zero. 
The maximum amount of GQD for the bipartite protocol of Sec.~\ref{sec: protocol} is ${\cal G}_{M_1M_2}\simeq0.2198$, {which is achieved when both carriers are projected upon states belonging to ``optimal subspaces" %}. The two carriers are not required to be measured in the same optimal basis in order to obtain the maximum amount of GQD; it is sufficient that they are each measured in any basis that is considered optimal. For the bipartite protocol, an optimal basis is that which have measurement outcomes 
identified by $\theta_{1,2}=0.9458,2.1958$ and arbitrary values of $\phi_{1,2}$.} %, corresponding to
%\begin{equation}\label{eq: optimal bases}
  %  \{\pm0.890\ket{0}+0.456e^{\ui\phi_i}\ket{1},\mp0.456\ket{0}+0.890e^{\ui\phi_i}\ket{1}\}\,,
%\end{equation}
%Again, the measurement outcomes as well as the values of $\phi_{1,2}$ do not affect the amount of distributed discord, although we note that more GQD is possible than for the bipartite definition of discord (see Sec.~\ref{sec: discord structure}).

In the remainder of this Section, we will present a study of the robustness of the protocol valid for a bipartite memory system. We address the effect of imperfections at the preparation and measurement stages of the protocol, quantifying the influences on the degree of achieved quantum discord.

\subsection{Measurement Precision}

{
Firstly, we consider how the measurement basis of the carriers affects the final discord between the memories. In Figs.~\ref{fig: heatmap}{\bf (a)}~and~\ref{fig: heatmap}{\bf (b)}, we present the behavior of the two asymmetric discord definitions against $\theta_{1,2}$. %These figures illustrate the asymmetric nature of these discord definitions, where we see that when the maximum of one definitions is reached, the other is exactly zero. 
In order to capture the quantum nature of all the quantum correlations, we also display our findings for the GQD in Fig.~\ref{fig: heatmap}{\bf (c)}. This figure appears somewhat as a superposition of Figs.~\ref{fig: heatmap}{\bf (a)}~and~\ref{fig: heatmap}{\bf (b)}, though not adding together linearly. The maximum values of the asymmetric heatmaps are also apparent on the GQD one, although we see that higher GQD values are possible in some places where both asymmetric discords are non-zero. A further discussion to this can be found in Sec.~\ref{sec: discord structure}. Beside the edges of Fig.~\ref{fig: heatmap}{\bf (c)} -- where, as discussed in the appendix, the distributed discord is null as the corresponding channel is both unital and semiclassical~\cite{StreltsovIncrease} -- no quantum correlations are shared for  $\theta_{1,2}=\pi/2$, for which %As one may expect, the discord is zero at the edges of the plot where the channel is semiclassical. The discord is non-zero everywhere else except for the measurement parameters $\theta_{1,2}=\pi/2$, which leaves 
the memories are left in the state $\rho_{M_1M_2}=(\ketbra{00}{00} + \ketbra{11}{11})/2$.
}

From this analysis, it is clear that imprecise measurement settings would result in degrees of quantum discord that could be significantly different from the maximum achievable as illustrated previously. %However, the channel is actually {not} semiclassical in this case, however a non-semiclassical channel does not imply that discord is generated (even if the channel is also non-unital).
In order to evaluate such effects quantitatively, we assume the parameters $\theta_{1,2}$ defining the basis upon which a measurement of the carriers' state should be performed to be random variables -- centred in the value that would allow us to achieve maximum discord -- and varying uniformly within a range of %We also consider how an imprecise measurement setting would affect the amount of distributed discord. Integrating the discord function with respect to the measurement parameters over a range of 
a chosen width. As a figure of merit, we evaluate the average degree of discord achieved by letting the parameters vary. For an imprecision range as large as $\pi/10$, though, a reduction of the maximum degree of GQD of only $~2\%$ is achieved, thus demonstrating a good robustness of the protocol.%  The average discord achieved within a range as large as $\pi/10$, $\pi/20$, and $\pi/30$ and then dividing by each of the ranges gives average GQD values of 0.2156, 0.2188 and 0.2193 respectively. Even when the measurement apparatus cannot guarantee precision, there is still very significant amounts of discord that can be distributed on average, showing that the protocol is robust to imperfections regarding the carrier measurements.

\subsection{Initial State of the Carriers}
\begin{figure}
   \centering
   \includegraphics[width=\linewidth]{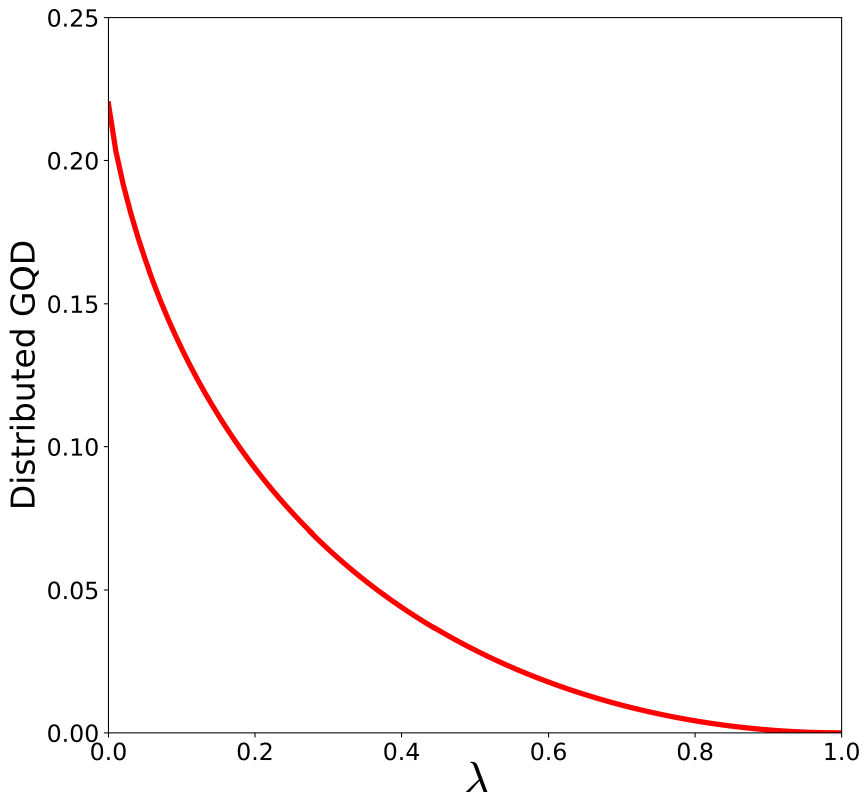}
   \caption{A study of the bipartite GQD distributed to the memories against the mixing parameter $\lambda$ in Eq.~\eqref{eq: carrier state mixed} and obtained assuming the optimal measurement bases for the carriers for any value of $\lambda$.}
   \label{fig: mixing parameter}
\end{figure}

Where the protocol is perhaps less robust, however, is with the initial state of the carriers. To investigate this, we consider the following convex combination of the desired carriers' state $\rho_{C_1C_2}$ and a state having no support on it, such as %$\sigma(\lambda)$ defined by mixing parameter $0\leq \lambda\leq 1$:
\begin{equation}\label{eq: carrier state mixed}
    \sigma(\lambda)=\left(1-{\lambda}\right)\rho_{C_1C_2} + \frac{\lambda}{2}\left(\ketbra{+-}{+-} + \ketbra{-+}{-+}\right)_{C_1C_2}
\end{equation}
with $\lambda\in[0,1]$. %For simplicity, we assume this state always has an equal contribution of both the $\ket{+-}$ and $\ket{-+}$ states. 
We then consider the amount of GQD that is obtained if we measure the carriers in the optimal basis (for the ideal case) with different values of the mixing parameter for the initial carrier state. The results are presented in  Fig.~\ref{fig: mixing parameter}, where it can be seen that the amount of GQD between the memories is fairly sensitive to mixing parameter $\lambda$, and it reaches zero for $\lambda=1$. We also calculate the average GQD distributed in the range $[0,0.1]$ such that the robustness is tested in the case where the fidelity of the initial carriers cannot be guaranteed, which gives $0.1687$: noticeably lower, yet still a significant amount of distributed GQD. 

Remarkably,  Fig~\ref{fig: mixing parameter} is identical to the graph obtained when maximising the GQD over $\theta_{1,2}$ at each value of $\lambda$, thus implying that the optimal bases for the ideal carrier state retain their optimality  for all states defined as in Eq.~\eqref{eq: carrier state mixed}. We investigate this further by generalising Eq.~\eqref{eq: carrier state mixed} to assume that the contributions of the $\ketbra{++}{++}$ and $\ketbra{+-}{+-}$ states in the mixture are not equal to the contributions of $\ketbra{--}{--}$ and $\ketbra{-+}{-+}$ states, respectively. {Similarly, we find that less initial classical correlations results in a lower degree of GQD. However, when the initial carrier state $\frac{1}{2}(\ketbra{+-}{+-}+\ketbra{-+}{-+})$ is used, the optimal amount of GQD generated is the same as when we use $\sigma(0)$ and for the same optimal carrier measurement basis.} This shows the robustness of the protocol to phase flips of the carrier qubits.

To conclude this part of our analysis, we now assess the interplay between purity and classical correlations in the initial carrier state, which we take as %The findings here suggest that either classical correlations, a certain degree of purity, or a combination of the two are required in the initial carrier state to distribute quantum correlations with this protocol, without any initial quantum correlations between the carriers. To give us an idea as to which of these is the true resource here for the discord distribution, 
\begin{equation}
    \rho_{C_1C_2}(\eta) =\left( \eta\ketbra{++}{++}+ (1-\eta)\ketbra{--}{--}\right)_{C_1C_2}
\end{equation}
with $\eta\in[0,1]$. Clearly, $\eta=1/2$ corresponds to the state in Eq. \eqref{eq: initial carriers}, while increasing (decreasing) the parameter towards 1 (0)  increases the purity of the state while simultaneously decreasing its classical correlations. Calculating the maximum possible distributed GQD as a function of $\eta$, we find -- quite expectedly -- that as $\eta\to0$ or $1$, no  GQD can be distributed. In general,  GQD decreases as $\eta$ deviates from 1/2 to disappear when such parameter takes its extremal values. 
This provides numerical evidence that classical correlations are the true resource of the protocol.
% Although this doesn't necessarily imply anything in general, it does provide numerical evidence that it is classical correlations that is the resource for GQD being distributed in this way, rather than the purity.

\subsection{Initial State of the Memories}
\begin{figure*}
    \centering
    \includegraphics[width=\linewidth]{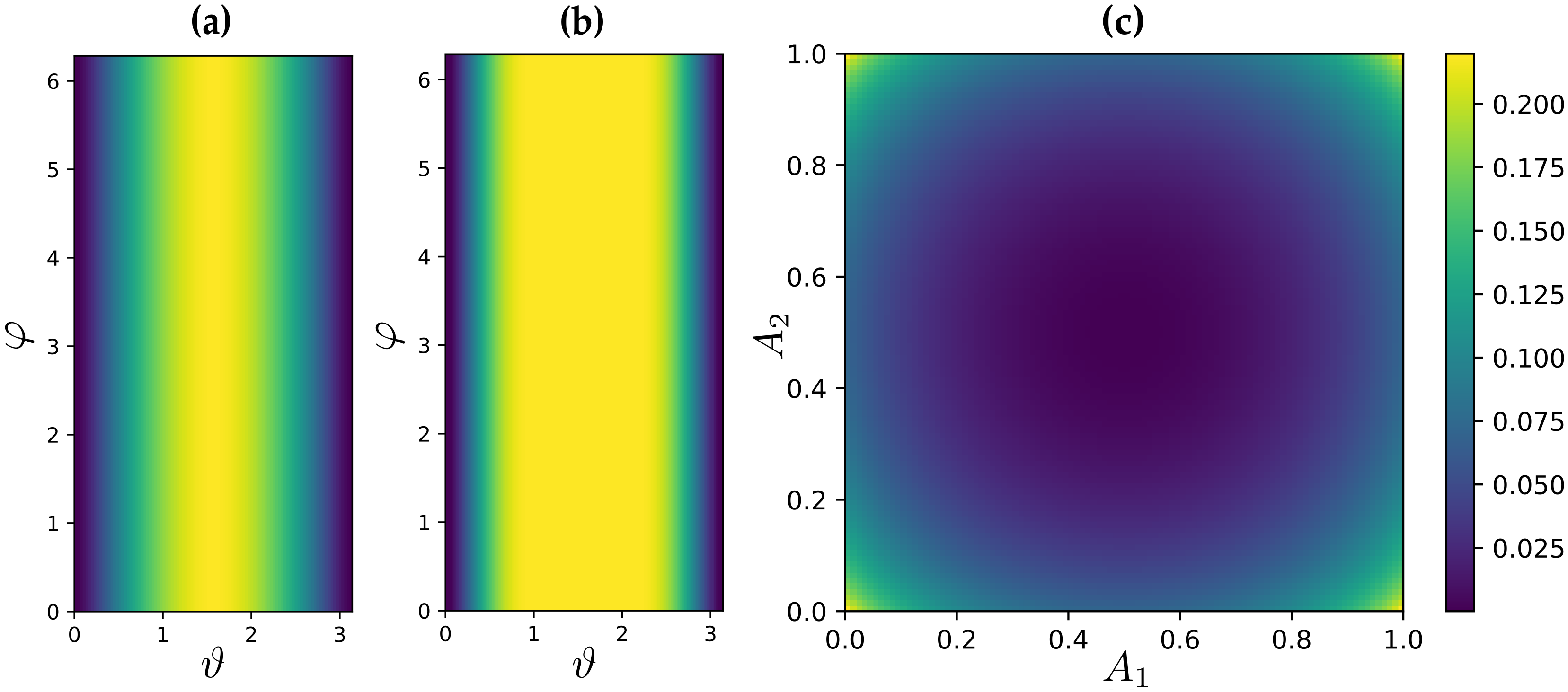}
    \caption{{The distributed GQD in the final $M_1$-$M_2$ compound as a function of variables which define the initial state of memories. We first consider memories to be in a pure product state defined by $\vartheta$ and $\varphi$; this is shown both for \textbf{(a)} the assumption that the carrier measurement bases used are characterized by $\theta_{1,2}=0.9458$ and $\phi_1=\phi_2=0$; and \textbf{(b)} optimization of the carrier measurements to show the largest amount of distributed GQD possible for each initial pure state of $M_2$. Note that heatmap \textbf{(a)} will change depending on the value of $\phi_2$, but will always be at a maximum at $\vartheta=\pi/2$. In \textbf{(c)} we display the GQD in the final $M_1$-$M_2$ compound as a function of the mixed-state parameters $A_1$ and $A_2$, where we have again assumed $\theta_{1,2}=0.9458$.}}
    \label{fig: pure memory same params}
\end{figure*}
We now test the robustness of the protocol with respect to the initial state of the memories by considering two scenarios: pure states and mixed states. In the former, we consider the memories to be prepared in a pure product state, but where the fidelity of $M_2$ with $\ket{+}_{M_2}$ is non unit. In the latter, we consider the case where the initial state of both memories may have been subject to some level of noise, so that both memories are in mixed states.

We start considering the initial state of $M_2$ %using the Bloch parameters $\vartheta\in[0,\pi]$ and $\varphi\in[0,2\pi)$ to define the pure state
\begin{equation}\label{eq: initial memory pure state}
    \ket{\psi(\vartheta,\varphi)}_{M_2} = \cos \frac{\vartheta}{2}\ket{0} + \sin\frac{\vartheta}{2}e^{\ui\varphi}\ket{1}\,.
\end{equation}
where we have introduced the Bloch parameters $\vartheta\in[0,\pi]$ and $\varphi\in[0,2\pi)$. 
We plot the GQD of the final state against $\vartheta$ and $\varphi$ in Figs.~\ref{fig: pure memory same params}{\bf (a)} and~\ref{fig: pure memory same params}{\bf (b)}. {We do not change the measurement basis at this point and instead measure the carriers in the same basis which maximizes discord when $\rho_{M_1 M_2} = \ketbra{++}{++}$.}
% \begin{figure}
%     \centering
%     \includegraphics[width=\linewidth]{memory_states_pure_figs.pdf}
%     \caption{\MP{the caption should be rewritten:} \AGH{The distributed GQD in the $M_1$-$M_2$ compound as a function of the pure state variables $\vartheta$ and $\varphi$. This is shown both for \textbf{(a)} the assumption that the carrier measurement bases used are characterised by $\theta_{1,2}=0.0458$ and $\phi_1=\phi_2=0$; and \textbf{(b)} optimisation of the carrier measurements to show the largest amount of distributed GQD possible for each initial pure state of $M_2$. Note that heatmap \textbf{(a)} will change depending on the value of $\phi_2$, but will always be at a maximum at $\vartheta=\pi/2$.}}
%     \label{fig: pure memory same params}
% \end{figure}
We find that, in analogy with the features highlighted previously on the carrier's measurement, %the basis upon which the initial state of $M_2$ is prepared is independent of the azimuthal angle, $\varphi$, and so 
the protocol is  robust to any azimuthal phase imprecision in the initial states of $M_2$. GQD is maximized for  $\vartheta=\pi/2$ and any choice for $\varphi$, so that states $\ket{\pm}$ and $\ket{\pm\ui}$ are all optimal. As noticed previously, GQD is weakly sensitive to small deviations from optimal initial states: for instance, the average GQD over a range symmetric around $\theta=\pi/2$ and of width $\pi/10$ {(with $\phi_{1,2}=0$)} %the the range $\vartheta\in[\pi(\frac{1}{2}-\frac{1}{20}),\pi(\frac{1}{2}+\frac{1}{20})]$ %(assuming the photon measurement basis remains the same as an optimal bases in Eq.~\eqref{eq: optimal bases}). This gives an average GQD of 
results in $\overline{\cal G}_{M_1M_2}= 0.2189$, which is a $\sim0.4\%$  reduction with respect to the ideal value. 

%However, we notice that the optimal bases identified for the ideal preparation of the memories are not necessarily optimal for all starting states of $M_2$, and a further exploration should be performed over the carrier measurements. In fact, the maximum GQD value of 0.2198 is achievable from $0.96\lesssim\vartheta\lesssim2.18$ so long as the carrier measurement basis is adjusted appropriately. \MP{can this paragraph be written a bit more clearly?}

{When a GQD value of $0.2198$ is achievable, the change of each carrier measurement basis (in order to become optimal) depends only on the initial pure state of the memory with which each carrier interacts. That is, in Fig~\ref{fig: pure memory same params}~\textbf{(b)}, only the measurement basis of $C_2$ need be adjusted in the range $0.96\lesssim\vartheta\lesssim2.18$. GQD here is, again, independent of the measurement outcome, however it does depend on the values of $\phi_2$ that define the state upon which $C_2$ is projected.} {For the cases where a maximum GQD value of $<0.2198$ is obtained in Fig~\ref{fig: pure memory same params}~\textbf{(b)}, the optimal measurement basis for $C_2$ appears to be $\{\ket{+\ui},\ket{-\ui}\}$, meanwhile the optimal measurement basis for $C_1$ depends on the initial pure state of $M_2$ (again, not depending on the measurement outcome).}

Let us now assume that both memories are initially in the following mixed state
\begin{equation}\label{eq: mixed state initial memories}
%\begin{split}
    \rho_{M_1M_2}(A_1,A_2){=} \bigotimes_{j=1,2}\big(A_j\ketbra{+}{+}+(1-A_j)\ketbra{-}{-}\big)_{M_j}%\\
%    &\otimes\big(B\ketbra{+}{+}+(1-B)\ketbra{-}{-}\big)_{M_2},
%\end{split}
\end{equation}
with $A_j\in[0,1]$. In Fig.~\ref{fig: pure memory same params}{\bf (c)} we plot the amount of GQD obtained when using an optimal photon basis against $A_{1,2}$,
% \begin{figure}
%     \centering
%     \includegraphics[width=\linewidth]{same_params_mixed_memories_heatmap.pdf}
%     \caption{Heatmap showing the distributed GQD of $M_1M_2$ against the mixed state parameters $\vartheta$ and $\varphi$ which define the initial state of $M_2$ in Eq.~\eqref{eq: mixed state initial memories}. The state is pure when $A_{1,2}=\{0,1\}$. In this plot, we have assumed the use of the optimal carrier measurement bases.}
%     \label{fig: same params mixed memories heatmap}
% \end{figure}
confirming that GQD is maximized by preparing the memories in pure states, in line with the expectations stemming from  the results reported in Figs.~\ref{fig: pure memory same params}{\bf (a)} and~\ref{fig: pure memory same params}{\bf (b)}. Again, Fig.~\ref{fig: pure memory same params}{\bf (c)} is identical to the case where we optimize over all carrier measurements for each plot point, solidifying the fact that the optimal bases are such not only in ideal conditions, but also for many non-ideal initial conditions that we consider in this paper. Thus,  the maximum amount of GQD possible for a given initial state could be obtained without the need to change the measurement basis. The only counterexample we find in this paper is for some initial pure states of the memories with a fidelity of less than unity with $\ket{+}$ or $\ket{-}$, as shown in Fig.~\ref{fig: pure memory same params}.

%To test how robust the GQD is to the memories being in a mixed state, we find the averages of GQD over certain intervals of $A$ and $B$. We integrate the GQD over the intervals $[0.9,1]$ and $[0.99,1]$ for both $A$ and $B$, giving average GQD values of 0.1359 and 0.2032 respectively. Integrating over just $B\in[0.9,1]$ and $B\in[0.99,1]$ (taking $A=1$) in order to gauge how much just one non-precise state preparation may affect GQD, an average GQD of 0.1692 and 0.2105 were found respectively.

We include an assessment of the performance of the protocol when the memories experience correlated dephasing noise as an appendix to this manuscript, detailing instances in which the protocol can actually generate \textit{more} quantum correlations as a result of such noise.

\section{Analysis of multipartite memory systems}\label{sec: system sizes}
We now address systems of memory systems of a growing size, thus addressing explicitly the distribution of multipartite quantum discord. To such end, we start considering the same protocol as in Sec.~\ref{sec: protocol}, but with $N$ carriers and  memories. The former are initially prepared in 
\begin{equation}
\begin{aligned}
    \rho_{\{C\}} = \frac{1}{2}(&\ketbra{++\cdots+}{++\cdots+}\\
    &+\ketbra{--\cdots-}{--\cdots-})_{C_1C_2\cdots C_N}\,,
\end{aligned}
\end{equation}
where $\{C\}$ stands for the collection of $N$ carrier systems. 
As for the memories, we assume them to be prepared in the $N$-party product state $\rho_{\{M\}}=\ket{++\cdots+}_{M_1M_2\cdots M_N}$. As in the bipartite case, each carrier $C_j$ and memory $M_j$ are subjected to a controlled-$Z$ interaction, with the former being measured in a suitable basis to establish GQD between the memories. In Table~\ref{tab: GQD} we report the distributed GQD  for  $N=2,.\dots,5$ resulting from an extensive numerical search for the best possible performance. The values presented in Table~\ref{tab: GQD} thus embody, at least, lower bounds to the maximum possible amount of distributed GQD to the memories \footnote{Higher-dimensional systems than 5 qubits were also attempted, however the simulations and optimizations were too computationally expensive.}. %These lower bounds, along with examples of bases which give these maxima (defined by measurement parameters $\theta_i$), are displayed in Table~\ref{tab: GQD}.
\begin{table}[b]
    \centering
    \begin{tabular}{c|c|c|c|c|c}
    \textbf{$N$} & \textbf{$\mathcal{G}_{M}$} & $\{\theta_i\}$ & $\mathcal{G}_W$ & $\mathcal{G}_\varepsilon$ & $\mathcal{G}_M/\mathcal{G}_\varepsilon$\\
    \hline
    \hline
    2 & 0.2198 & 0.9458 & 1.0000 & 0.4124 & 0.5330\\
    \hline
    3 & 0.4694 & 0.9131 & 1.5850 & 0.8070 & 0.5817\\
    \hline
    4 & 0.7040 & 0.8775 & 2.0000 & 1.1554 & 0.6093\\
    \hline
    5 & 0.9338 & 0.8533 & 2.3219 & 1.3749 & 0.6792\\
    \hline
    \end{tabular}
    \caption{We report the values of $\mathcal{G}_{M} = \max_{\{\theta_i\}}[\mathcal{G}(\rho_{\{M\}}^\prime)]$, which represents a lower bound for the maximum amount of GQD possible in the  final memory state $\rho_{\{M\}}^\prime$. We also report the corresponding measurement parameters $\{\theta_i\}$ to identify exemplary optimal bases for all carriers to be measured in. In order to provide meaningful benchmarks, we display $\mathcal{G}_W=\mathcal{G}(\ketbra{W_N}{W_N})$ and $\mathcal{G}_{\varepsilon}=\mathcal{G}(\rho_{W,N}(\varepsilon))$, the latter being the GQD of the state which has the same level of mixedness as each optimized $\rho_{\{M\}}^\prime$. The ratios $\mathcal{G}_M/\mathcal{G}_\varepsilon$ give a simple figure of merit to track the scaling of the distributed GQD with the size of the system.}
    \label{tab: GQD}
\end{table}
We remark that, in line with the bipartite case, for $N\ge3$ the same measurement basis will be needed for all the carriers, while the degree of distributed discord does not depend on the specific outcome of such measurements. %  of the the measurement basis remains the same for all carriers, with the measurement outcomes again not affecting the amount of distributed GQD.
\begin{table*}[t]
\centerline{%
    \begin{tabular}{c"c|c|c|c|c|c|c|c|c|c}
         & \multicolumn{3}{c|}{\kern-0.6em 12} & \multicolumn{3}{c|}{23} & \multicolumn{3}{c|}{13} & 123 \\
    $\{i\}$ int. & Cmpd. & $D_{1\vert 2}$ & $D_{2\vert 1}$ & Cmpd. & $D_{2\vert 3}$ & $D_{3\vert 2}$ & Cmpd. & $D_{1\vert 3}$ & $D_{3\vert 1}$ & Max. GQD \\
       \hline
       \hline
 %       None & $C_1C_2$ & \xmark & \xmark & $C_2C_3$ & \xmark &  \xmark & $C_1C_3$ & \xmark & \xmark & 0 \\
  %      \hline
         $\{1\}$ & $M_1C_2$ & \xmark & \cmark & $C_2C_3$ & \xmark & \xmark & $M_1C_3$ & \xmark & \cmark & 0.2018 \\
        \hline
         $\{1,2\}$ & $M_1M_2$ & \cmark & \cmark & $M_2C_3$ & \xmark & \cmark & $M_1C_3$ & \xmark & \cmark & 0.4036 \\
        \hline
         $\{1,2,3\}$ & $M_1M_2$ & \cmark & \cmark & $M_2M_3$ & \cmark & \cmark & $M_1M_3$ & \cmark & \cmark & 0.4694 \\
        \hline
    \end{tabular}}
% \vspace*{0.3 cm}
% \centerline{%
%     \begin{tabular}{c"c|c|c|c|c|c|c|c|c|c|c|c|c|}
%          & \multicolumn{2}{c|}{\kern-0.6em 12} & \multicolumn{2}{c|}{23} & \multicolumn{2}{c|}{13} & \multicolumn{2}{c|}{14} & \multicolumn{2}{c|}{24} & \multicolumn{2}{c|}{34} & 123 \\
%        $\{i\}$ int.  & $D_{1|2}$ & $D_{2|1}$  & $D_{2|3}$ & $D_{3|2}$  & $D_{1|3}$ & $D_{3|1}$ & $D_{1|4}$ & $D_{4|1}$ & $D_{2|4}$ & $D_{4|2}$ & $D_{3|4}$ & $D_{4|3}$ & Max. GQD \\
%        \thickhline
%         None & \xmark & \xmark  & \xmark &  \xmark  & \xmark & \xmark & \xmark & \xmark & \xmark & \xmark & \xmark & \xmark & 0 \\
%         \hline
%          $\{1\}$ & \xmark & \cmark & \xmark & \xmark & \xmark & \cmark & \xmark & \cmark & \xmark & \xmark & \xmark & \xmark & 0.20175 \\
%         \hline
%          $\{1,2\}$ & \cmark & \cmark & \xmark & \cmark & \xmark & \cmark & \xmark & \cmark & \xmark & \cmark & \xmark & \xmark & 0.40350 \\
%         \hline
%          $\{1,2,3\}$ & \cmark & \cmark & \cmark & \cmark & \cmark & \cmark & \xmark & \cmark & \xmark & \cmark & \xmark & \cmark & 0.60526 \\
%         \hline
%         $\{1,2,3,4\}$ & \cmark & \cmark & \cmark & \cmark & \cmark & \cmark & \cmark & \cmark & \cmark & \cmark & \cmark & \cmark & 0.70402 \\
%         \hline
%     \end{tabular}}
    \caption{Table displaying the discord structure of three different three-qubit discord distribution protocols, depending on which of the compounds (cmpds.) $\{C_i$-$M_i\}$ involve an interaction and measurement (int.). A \xmark\  indicates no discord is possible, meanwhile a \cmark\ indicates that the discord is non-zero in general (up to a maximum of 0.20175). Also displayed are the maximum possible values of the GQD of the total final state of each (assuming any measured carriers and unused memories are traced out). We note that the maximum GQD of the state with two interactions involved has a GQD which is exactly double the maximum asymmetric discord value of any resulting bipartite compound.}
    \label{tab: discord structure}
\end{table*}

To provide a benchmark for the significance of the amount of distributed GQD for each system size, we also provide the amount of GQD present in the $W$ state 
\begin{equation}
    \ket{W_N} = \frac{1}{\sqrt{N}}\ket{N-1,1}\,,
\end{equation}
where $\ket{N-1,1}$ represents the totally symmetric state containing $N-1$ zeros and 1 one~\cite{Dur2000}.
We stress that these states are not necessarily those with the largest amount of discord for each system size~\footnote{Unlike the bipartite case, some multipartite \textit{pure} states -- including $\ket{W_3}$ -- may possess other quantum correlations in addition to entanglement, which have been dubbed \textit{quantum dissonance}~\cite{ModiRelative}.} -- despite being maximally entangled -- and serve only as a reference.
Along with this, we consider the GQD of the Werner state $\rho_{W,N}$ for each $N$ which has the same level of mixedness as the resulting state of the memories at the end of each protocol. We define $\rho_{W,N}$ as %the $\ket{W_N}$ state subjected to some level of white noise, dictated by some mixing parameter $\varepsilon$, i.e.
\begin{equation}
    \rho_{W,N}(\varepsilon) = (1-\varepsilon)\ketbra{W_N}{W_N} + \frac{\varepsilon}{2^N}\openone~~~\left(\varepsilon\in[0,1]\right)\,.
\end{equation}
The ratio between the lower bound to the GQD of the memories achieved through our protocol and the corresponding quantity carried by the Werner state of the same degree of mixedness grows with $N$ for all the cases numerically addressed here. Although we cannot confidently extrapolate a general behavior, this suggests that, not only does  GQD increase with $N$, but also the amount of GQD with respect to allowed amount by system size improves. To give us an idea of the maximum amount of GQD for a given $N$, we use the few plot points we have to calculate the linear regression between the two: this gives us an approximate GQD of $0.238N-0.250$ for $N$ qubits. {In Sec.~\ref{sec: discord structure} we put forward some arguments against the accuracy of such (intuitively reasonable) linear model.}% -- noting again that we cannot consider this to be accurate, and serves only to give us an idea of what the relation could be.

%\section{Practical implementation of the protocol}\label{sec: implementation}

To conclude this section, we briefly assess the performance of the protocol when it is performed using two carrier qubits which are in the tripartite GHZ state $\rho_{C_1C_2C_3} = \ketbra{\mathrm{GHZ}_3}{\mathrm{GHZ}_3}\text{, where }\ket{\mathrm{GHZ}_3} = \frac{1}{\sqrt{2}}(\ket{+++}+\ket{---})_{C_1C_2C_3}$. We do this since tracing out one of the carrier qubits from this state leaves the remaining two in the classically correlated state in Eq.~\eqref{eq: initial carriers}. Assuming the protocol is performed for the memories $M_1$ and $M_2$ using carriers $C_1$ and $C_2$, we verify that the maximum GQD in the final state of the compound $M_1$-$M_2$ cannot be increased beyond the usual maximum, however a GQD increase from 1 (in the initial GHZ carrier state) to 1.2926 is possible for the state comprising $M_1$-$M_2$-$C_3$. Any entanglement present in the state is lost when $C_3$ is traced out; this being true at any stage of the protocol.

\section{Discord Structure of the Final State}\label{sec: discord structure}
\begin{figure*}[t]
    \centering
    \includegraphics[width=\linewidth]{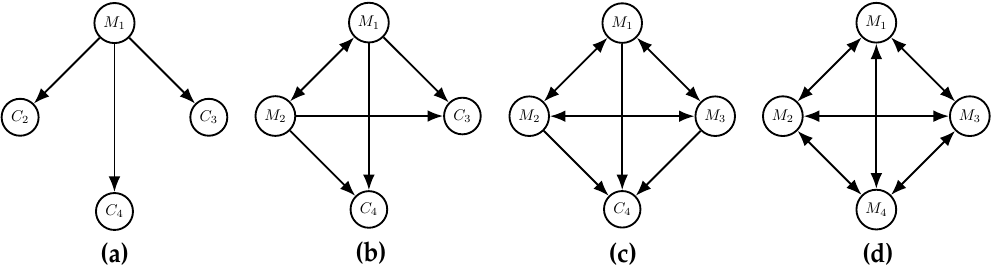}
    \caption{{Graphical representation of the discord structure of the final state of four differing discord distribution protocols with four qubits in the final state, based on the discord structure graphs shown in Fig.~\ref{fig: discord structure}. These four protocols involve a growing number of interactions going from one to four (moving from panel \textbf{(a)} to \textbf{(d)}). All  interactions occur between carriers and memories. Where an interaction (and measurement of carrier) takes place, a memory $M_i$ is used in the final state with $C_i$ being traced out. Where no interaction occurs, the unused memory is instead traced out in place of the carrier.}}
    \label{fig: 4 discord structure}
\end{figure*}

We now conduct a deeper study of the discord structure of the prepared bipartite or multipartite discordant states.
%As shown in Secs.~\ref{sec: protocol} and \ref{sec: robustness}, the maximum amount of asymmetric discord possible from the bipartite protocol is $D_{M_1\vert M_2}\simeq0.2018$, while the corresponding maximum GQD is ${\cal G}_{M_1M_2}\simeq0.2198$ resulting from different carrier measurement bases.
As we have already pointed out, the definition of discord in Eq.~\eqref{defdiscord} is inherently asymmetric and, in fact, we have $D_{M_2\vert M_1}=0$ {when $D_{M_1\vert M_2}$ is maximized}. %This occurs for the carrier measurement parameters $\theta_1=\pi/2,\theta_2=\pi/4$; if these are switched to $\theta_1=\pi/4,\theta_2=\pi/2$ then we instead find $D_{M_1\vert M_2}=0$ and $D_{M_2\vert M_1}\simeq0.2018$. Meanwhile, 
On the other hand, it is interesting to analyze the state which achieves the maximum degree of GQD allowed by our protocol. As stated in Sec.~\ref{sec: robustness}, this is achieved for carrier measurements with by $\theta_1=\theta_2=0.9458$. A calculation of the degree of asymmetric discord for such state leads to $D_{M_1\vert M_2}=D_{M_1\vert M_2}=0.1498$, showing the quantum-quantum nature of such resource and thus an inherently different sharing structure of quantum correlations within it than the one achieved by using the measurement settings maximizing $D_{M_1\vert M_2}$.

%\subsection{Higher-dimensional Multipartite Protocols}
{A similar line of thought can be pursued in the case of a multipartite system. We first consider the tripartite protocol, which we attack by implementing three different protocols, each identified by how many $C_i$-$M_i$ compounds undergo controlled-$Z$ interactions (followed by a measurement of $C_i$). For any $C_i$-$M_i$ compound that does not experience such operations, we use $C_i$ in the final state instead of $M_i$, which we trace out. We do this so as to pin-point the dependence of both the structure and quantity of the quantum correlations on which local operations are performed.  %but where the local operations are not necessarily performed on all $C_iM_i$ compound states (where $C_i$ represents the $i$th carrier and $M$ the $i$th memory). The state is symmetric so we need not consider every single permutation of operations performed/not performed; we instead consider the case of local operations being performed on 1, 2, and all 3 compound states and assessing the final discord structure. 
As the initial states of the system are invariant under particle exchange, we need not consider every permutation of operations performed, rather just the number of operations performed.  %We start with no local operation is performed, then $C_i$ is used in the final state instead of $M_i$. We find the bipartite discord in each of the bipartite compounds of the final state and display our results in 
Table~\ref{tab: discord structure} presents the results of our quantitative analysis.}

We find that $D_{X_i\vert Y_j}=0$ when $Y=C$ and $D_{X_i\vert Y_j}\neq0$ when $Y=M$, for $X,Y\in\{C,M\}$ and $i\neq j$ are the labels identifying each carrier-memory compound ($i,j\in\{1,2,3\}$ for the tripartite protocol). Thus, if a carrier $C_i$ interacts with a memory $M_i$ before being measured, this, in general, creates asymmetric discord between $M_i$ and the other states. Additionally, if, for $j\neq i$, the carrier $C_j$ interacts with the memory $M_j$ before being measured (and as long as the interaction happens before any other carrier measurements are made), this will also, in general, create asymmetric discord between $M_j$ and $M_i$. Although, in general, discord is non-zero in both 'directions', we still call it asymmetric as $0<D_{M_i\vert M_j}\neq D_{M_j\vert M_i}>0$. To illustrate how this trend continues for higher-dimensional protocols more succinctly, we display the discord structure of the four possible ($N=4$)-qubit protocols graphically in Fig.~\ref{fig: 4 discord structure}. Although we have not pursued this investigation further, we expect this trend to continue for all system sizes.

The maximum amount of asymmetric discord in any bipartite compound with non-zero discord is 0.2018, i.e. the same amount possible in the bipartite protocol. We see that, in the tripartite protocol, the maximum GQD in the resulting state when two sets of local operations occur is exactly double that value, meanwhile when all three involve local operations then it is slightly more than double. We find this pattern to similarly continue for the four qubit protocol: when three out of the four subsystems involve local operations, the maximum GQD is exactly triple the value of 0.2018, where again the maximum GQD when all memories are interacted with {(and carriers are subsequently measured)} is slightly more than triple. This suggests that the {linear relation between the maximum GQD and the system size given earlier in this section}
%in Sec.~\ref{sec: system sizes}
could be replaced by the more accurate one
\begin{equation}
    \mathcal{M}(N) = 0.2018(N-1) + \xi(N)\,,
\end{equation}
where $\xi$ is some {non-linear function} determining how much 'additional' GQD is possible for each system size. We conjecture the form of this function by assuming an exponential fit, predicting
%A non-linear fit gives us 
%We conjecture the form of this function by fitting the additional GQD -- i.e., the maximum GQD minus $0.2018(N-1)$ -- against the system size $N$ and assuming an exponential fit, predicting
\begin{equation}
    \xi(N) \simeq -0.3320e^{-0.2863N}+0.2056
     \qquad(N>1)
    \,.
\end{equation}

\section{Conclusions and Outlook} \label{sec: conclusions}
We have presented a protocol by which significant amounts of quantum correlations can be distributed to quantum memories through a classically correlated resource state. The amount of discord distributed depends only on the measurement basis, regardless of the measurement outcomes. We have also provided numerical evidence of scalability for a desired number of quantum memories, which may all share large amounts of discord. In addition, the robustness of the protocol has been evaluated, revealing an inherent robustness to imprecise measurement settings. %although more sensitive to the fidelity of the initial state.
Recent work has also shown that, with larger amounts of classical correlation in the resource state, it is also possible to generate other types of discordant states through local operations such as Werner states \cite{Gokhan}. We explore how our protocol may be adapted for this possibility with a correlated noise channel applied to the memories in App.~II.

In future work, one may consider how to use the distributed discord network in a practical way, and the protocol here is yet to stand up to experimental scrutiny. Moreover, the possibility of using weakly-entangled initial states to enable higher amounts of distributed discord can be explored. The true usefulness of a discord distribution scheme like the one detailed here remains to be seen, owing to the fact that discord may still have unexplored potential as a resource in quantum information processing, even if only as a vehicle for establishing entanglement. While discord is the resource for entanglement distribution, classical correlation is a resource for discord distribution.

{The protocol proposed here is versed to implementation in at least two mainstream platforms, namely cavity quantum electrodynamics (QED) \cite{Walther,Monroe} and its circuital version \cite{Blais}. In both platforms, which have enjoyed significant success in demonstrating the power of measurement-enhanced coherent information processing \cite{Micuda_2020,Starek_2021,Zanin_2022}, the carrier states could be encoded in the polarisation degree of freedom of radiation-based qubits (microwave signals in the case of circuit  QED). The memories, instead, would be encoded in neutral atoms trapped in optical cavities or superconducting transmon qubits embedded in stripline resonators. In both setups, the controlled-$Z$ operations required can be realized from light-matter interactions formally describable in terms of the effective operations reported in Ref.~\cite{DuanPhotonSpin}, even if the probability of such operations failing nears unity \cite{DuanProb}.
This potential experimental setting would provide avenues for space-to-ground (or ground-to-ground) discord distribution, with recent work exploring zero-added loss photon sources \cite{Chen}.
% These studies focus on mapping entangled states of photons onto memories, however it must be noted that entanglement is a very fragile resource, unlike classical correlations which we use as a catalyst for quantum correlations.
%We emphasise this as an advantage here 
Using our protocol, the quantum correlations themselves would not be subject to any noise that the photons are subject to when travelling from the source to the labs, since these correlations would be created locally from the labs. While this brings an advantage, one must be careful in ensuring that the classical correlations remain robust to the environment, and future works may consider the robustness of this protocol to noise models. We also consider this setup to be ideal for distributed quantum computation, since non-local (entangling) gates can be performed on two remote atoms trapped in different optical cavities \cite{AtomGates}.}

The data generated as part of this work are available from Ref.~\cite{Zenodo} and, upon reasonable requests, from the authors.

\acknowledgements
 
We acknowledge support by
the European Union’s Horizon Europe EIC-Pathfinder
project QuCoM (101046973), the Royal Society Wolfson Fellowship (RSWF/R3/183013), the UK EPSRC
(EP/T028424/1), the Department for the Economy
Northern Ireland under the US-Ireland R\&D Partnership Programme, the ``Italian National Quantum Science and Technology Institute (NQSTI)" (PE0000023) - SPOKE 2 through project ASpEQCt, the “National Centre for HPC, Big Data and Quantum Computing (HPC)” (CN00000013) – SPOKE 10 through project HyQELM, and the EU Horizon Europe EIC Pathfinder project QuCoM (GA no.~10032223). 
 
\bibliography{refs.bib}

\section*{Appendix I: Critical analysis of the bipartite protocol}

%It may go against intuition that quantum correlations can be distributed between $M_1$ and $M_2$ in this way, since all the operations performed are local with respect to the $M_1:M_2$ bipartition.
 It has been shown that the action of certain local quantum channels on bipartite states {could} generate discord~\cite{StreltsovIncrease,CiccarelloChannels}. Examples of such a peculiar effect, which puts discord at variance with entanglement for which such a result is not possible, have been shown in Refs.~\cite{CampbellProp,Ciccarello,Lanyon}. The class of quantum channels that are able to generate discord locally are those that are neither \textit{semiclassical} nor \textit{unital}, as reported by Ref.~\cite{StreltsovIncrease}.
 
A semiclassical channel $\Lambda_{\mathrm{sc}}$ is one that maps all input states $\rho$ onto output ones $\Lambda_{\mathrm{sc}}(\rho)$ that are diagonal in the same basis. When considered against the state achieved through our protocol applied to the to the initial state $\rho_{M_1 M_2}$, we see that the final state is diagonal in the $\{\ket{++},\ket{+-},\ket{-+},\ket{--}\}$ basis only when $\theta_1=\{0,\pi\}$ and/or $\theta_2=\{0,\pi\}$ (i.e., when either carrier is measured in the computational basis). We have further found that these parameters will result in a semiclassical channel for any other classical input state {of the memories}. In all such instances, the resulting discord is zero. This is not implied if the initial state of the memories is discordant; rather, it is implied that the resulting \textit{increase} of discord will always be zero. For all other values of $\theta_{1,2}$, the channel entailed by our protocol may, in general, not be semiclassical.

Meanwhile, a unital channel $\Lambda_{\mathrm{un}}$ is one that maps the maximally mixed state ${\openone}/{d}$ of a $d$-dimensional system to itself, i.e.
\begin{equation}
    \Lambda_{\mathrm{un}}\left(\frac{1}{d}\openone\right) = \frac{1}{d}\openone.
\end{equation}
%with $\openone$ the $d\times d$ identity matrix.
When applied to the initial state $\rho_{M_1M_2}=\frac{1}{4}\openone$ of the memories, our protocol returns the  measurement-dependent state %that is returned by the channel is
\begin{equation}
\rho_{M_1M_2} = \frac{1}{4}\openone + \frac{1}{4}(\cos \phi_1 \cos \phi_2 \sin \theta_1 \sin \theta_2) Z_{M_1}\oplus\left(-Z_{M_2}\right),
\end{equation}
showing that, in general, the channel is not unital. However, the channel \textit{is} unital for one (or a combination) of the following cases: $\theta_{1,2}=0,\pi$ and $\phi_{1,2}=\pi/2,3\pi/2$. Since the amount of discord is independent of the values of $\phi_{1,2}$, we see that we can generate the maximum amount of discord here even from a unital channel, for example with the parameters $\theta_1=\pi/2$, $\theta_2=\pi/4$, $\phi_1=\pi/2$, $\phi_2=0$.

Despite the protocol resulting in a unital effective channel for the memories, our scheme is not in contradiction with the findings of Ref.~\cite{StreltsovIncrease} due to the initial state of the carriers. The (classical) correlations initially shared by $C_1$ and $C_2$ in Eq.~\eqref{eq: initial carriers} prevent the resulting quantum channel affecting the memories to be factorized, thus is not {\it mathematically local}. More formally, despite the local nature of the operations performed on the memories, the initial carrier correlations results in a channel $\Lambda_{M_1M_2}(\rho_{M_1M_2})\neq\Lambda_{M_1}(\rho_{M_1})\otimes\Lambda_{M_2}(\rho_{M_2})$ with $\rho_{M_j}=\Tr_{k\neq j}\rho_{M_1M_2}=\ketbra{+}{+}_{M_j}~(j,k=1,2)$ in light of Eq.~\eqref{eq: initial carriers}. %where it is stated that local unital channels cannot create discord if the channel is applied to just one of the qubits in the multiqubit system. Here, the channel is applied to more than one qubit and, although the operations performed are local with respect to the $M_1M_2$ compound, the channel is not \textit{mathematically} local, as the Kraus operators $\{K_{i}^{M}\}$ cannot be written in the form of $K_{i}^{M_1M_2}=K_i^{M_1}\otimes K_i^{M_2}$. This is likely due to the fact that the measurement of each carrier is required to be performed after both \textsc{cphase} interactions have been realised; if one of the carriers is measured before the other interaction has occurred, then this is not represented by the protocol channel we find here.

This point could be further corroborated by studying the performance of the protocol resulting from the use of just one memory, say $M_2$, initially in state $\ket{+}_{M_2}$. A controlled-$Z$ gate is applied to the joint state of $C_2$ and $M_2$ before $C_2$ is measured, while no action is taken on $C_1$. {In this case, any discord produced will be asymmetric and we generate classical-quantum states instead of quantum-quantum states}. For $\theta_2=\{\pi/4,3\pi/4\}$, the corresponding state of the $C_1$-$M_2$ compound carries the same amount of discord of 0.2018 found in the two-memory configuration. To test unitality, we act with this protocol on the initial state $\rho_{C_1C_2}\otimes \openone_{M_2}/2$, meaning the initial state of $C_1M_2$ (as well as $C_2M_2$) is maximally mixed. The corresponding final state of $C_1M_2$ is $\openone_{C_1M_2}/4$ for one of the choices $\theta_2 = \{0,\pi\}\text{ , } \phi_2 = \{\pi/2,3\pi/2\}$ being satisfied. However, the amount of discord seeded in the state of such compounds is again independent of $\phi_2$, thus implying that the  channel is discord-generating when unital, despite the locality of the considered operations. Nonetheless, we conclude that, since the initial state cannot be written as a tensor product of the form $\rho_{C_1M_2}\otimes\rho_{C_2}$ due to the initial classical correlations between $C_1$ and $C_2$, the statement in Ref.~\cite{StreltsovIncrease} again does not apply here. %The consequence of this is that the operators acting on each pure state in the mixture are different, depending on the pure state itself. Therefore, the operators are not Kraus operators and it is not a quantum channel like that assumed in Streltsov's paper, so the theorem does not apply here either.

\section*{Appendix II: Performance of the bipartite protocol under correlated dephasing noise}
\begin{figure*}
    \centering
    \includegraphics[width=0.9\linewidth]{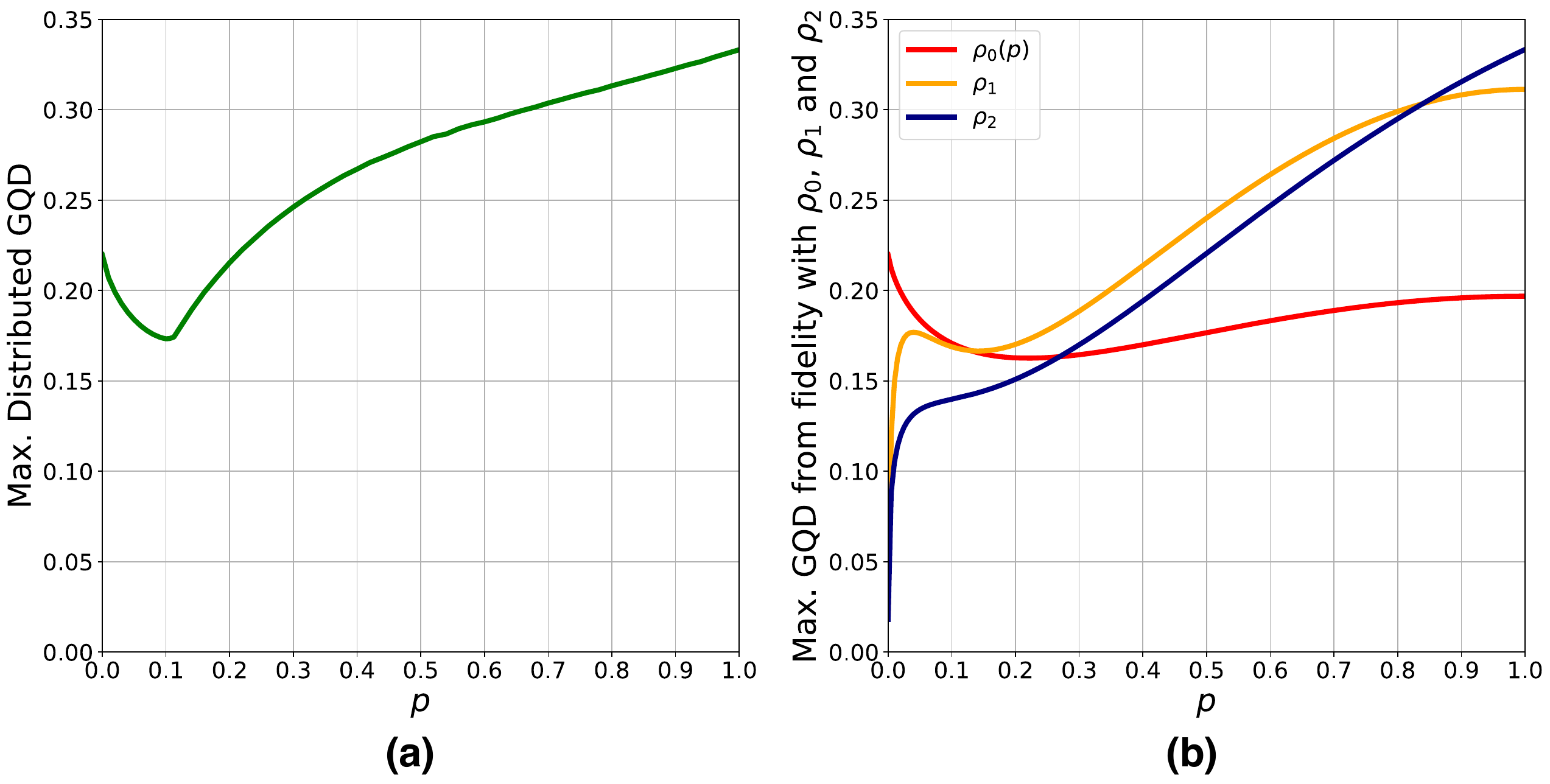}
    \caption{\textbf{(a)} The amount of GQD in the final memory state, maximized over the carrier measurement parameters, plotted as a function of the strength $p$ of a fully correlated dephasing channel acting on the initial state of the memories $M_1$ and $M_2$. \textbf{(b)} GQD of $\rho_0(p)$ along with the maximum amount of GQD possible (as a result of the protocol) from fidelity with the states $\rho_1$ and $\rho_2$, optimizing fidelity over carrier measurements.}
    \label{fig: appendix plots 1}
\end{figure*}
% \begin{figure}[b]
%     \centering
%     \includegraphics[width=\linewidth]{optimised_GQD_correlated_dephasing.pdf}
%     \caption{The amount of GQD in the final memory state, maximized over the carrier measurement parameters, plotted as a function of the strength $p$ of a fully correlated dephasing channel acting on the initial state of the memories $M_1$ and $M_2$.}
%     \label{fig: optimised correlated noise}
% \end{figure}
One of the motivations of our protocol is the ability to generate discord between two quantum memories via bi-local operations only, without and direct communication between said memories. With that being said, we also wish to explore how our protocol may perform under correlated noise being applied to the memories. We model such noise  using the following Kraus operators
\begin{align}
    K_1(p) &= \sqrt{1-\frac{p}{2}}(\Id_{M_1}\otimes\Id_{M_2})\\
    K_2(p, \mu) &= \sqrt{\frac{p}{2}(1-\mu)}(Z_{M_1}\otimes\Id_{M_2})\\
    K_3(p, \mu) &= \sqrt{\frac{p}{2}(1-\mu)}(\Id_{M_1}\otimes Z_{M_2})\\
    K_4(p,\mu) &= \sqrt{\frac{p}{2}\mu}(Z_{M_1}\otimes Z_{M_2})\,,
\end{align}
characterized by the strength of the noise $0\leq p \leq 1$ and the strength of the correlation $0\leq \mu\leq1$. We assume the extremal case $\mu=1$ in order to assess the performance of the protocol with as much classical correlation in the initial $C_1M_1:C_2M_2$ bipartition as possible. We compute the maximum GQD (with respect to the measurement parameters) for each value of $p$ and display our results in Fig.~\ref{fig: appendix plots 1}~\textbf{(a)}.
After an initial drop in the maximum amount of GQD, we actually find, rather counterintuitively, that more GQD is possible under correlated dephasing noise for strengths $p\gtrapprox0.21$ than for when there is no noise at all, obtaining a maximum GQD of $\frac{1}{3}$. for $p=1$. This rise in maximum GQD may not be completely unexpected, since this type of noise channel will increase the classical correlation in the previously-uncorrelated state. In fact, for $p=1$ the state in Eq~\eqref{eq: initial carriers} is now also the initial state of the memories, meaning twice the initial classical correlations in $C_1M_1:C_2M_2$. Nevertheless, what is especially peculiar about Fig.~\ref{fig: appendix plots 1}~\textbf{(a)} is the fall before the sudden rise at around $p\approx0.11$.
\begin{figure*}
    \centering
    \includegraphics[width=0.9\linewidth]{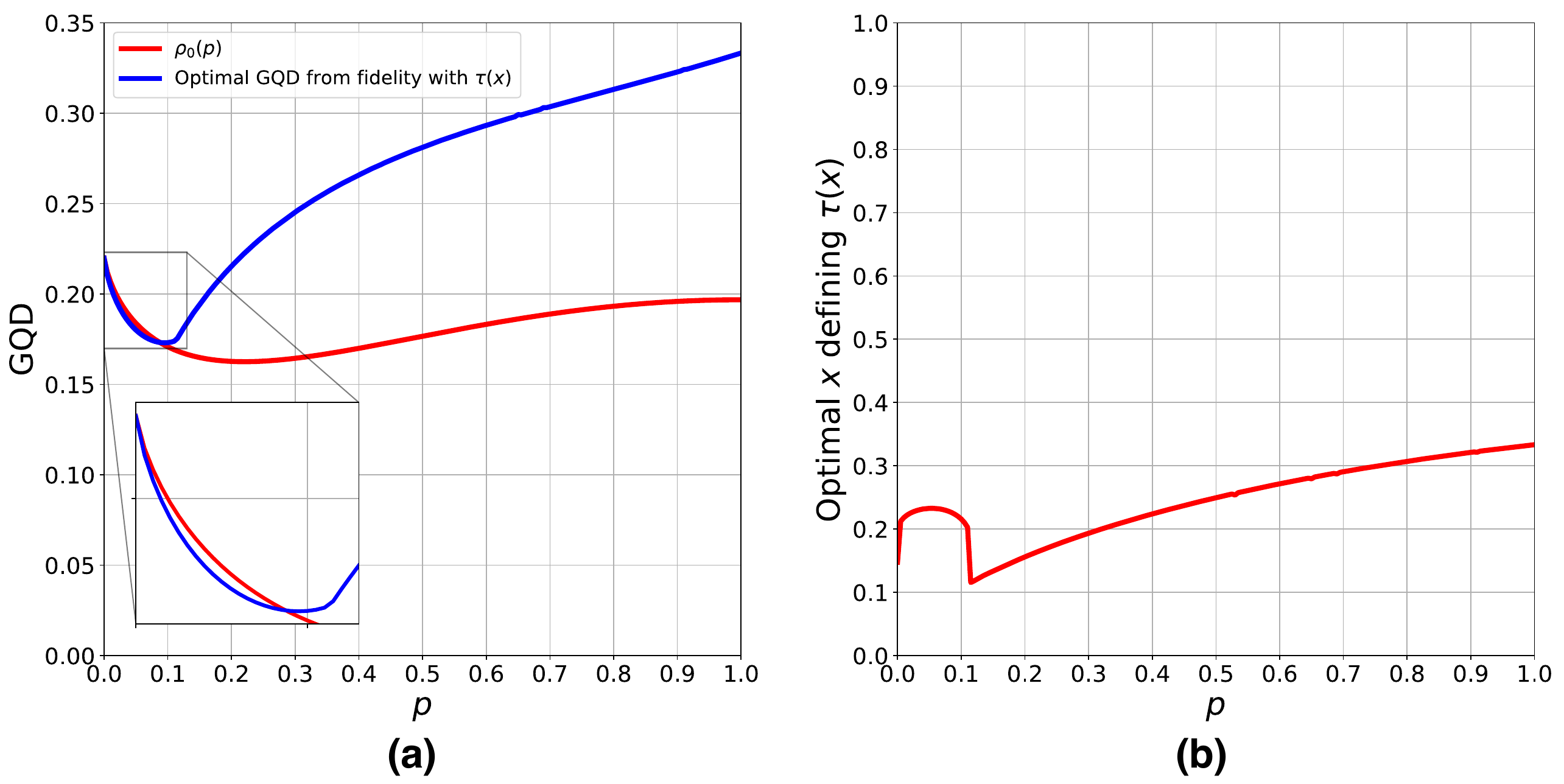}
    \caption{\textbf{(a)} GQD of $\rho_0(p)$ along with the maximum amount of GQD possible (with respect to the carrier measurements) when searching the maximum fidelity values with the $\tau(x)$ states (with respect to $x$). A zoomed region is shown for $0\leq p\leq0.13$. \textbf{(b)} Optimal $x$ values plotted against $p$. Each optimal $x$ defines the state $\tau(x)$ with which a high fidelity results in the most GQD for each $p$-value.}
    \label{fig: appendix plots 2}
\end{figure*}

To explain this behavior, we consider discordant states of different structures with which the final state of the memories may have a high fidelity with. Such states may include Werner states \cite{Gokhan} and mixtures of Bell states, so we consider the following three types of discordant states:
\begin{enumerate}
    \item For each $p$, the state resulting from the carrier measurement parameters being taken as the same used in the noiseless ($p=0$) case, assuming $\theta_1=\theta_2=0.9458$ and $\phi_1=\phi_2=0$:
    \begin{equation}
        \rho_0(p) = \rho_{M_1M_2}^\prime (p) \,.
    \end{equation}
    \item The following Bell state mixture:
    \begin{equation}
        \rho_1 = \frac{1}{2}\ketbra{\Psi^+}{\Psi^+} + \frac{1}{4}\left(\ketbra{\Phi^+}{\Phi^+} + \ketbra{\Phi^-}{\Phi^-}\right)\,.
    \end{equation}
    This state has a GQD of 0.3113 and is almost entangled; if it parameterised as
    \begin{equation}
        \tau(x)=x\ketbra{\Psi^+}{\Psi^+} + \frac{(1-x)}{2}\left(\ketbra{\Phi^+}{\Phi^+} + \ketbra{\Phi^-}{\Phi^-}\right)
    \end{equation}
    with $x\in[0,1]$, then it is entangled for $x>\frac{1}{2}$.
    \item The following Bell state mixture:
    \begin{equation}
        \rho_2 = \frac{1}{3}\left(\ketbra{\Psi^+}{\Psi^+} + \ketbra{\Phi^+}{\Phi^+} + \ketbra{\Phi^-}{\Phi^-}\right)\,.
    \end{equation}
    This state is equivalent to $\tau(\frac{1}{3})$ and has a GQD of exactly $\frac{1}{3}$, so it is further from being an entangled state than $\tau(\frac{1}{2})$ despite being more discordant.
\end{enumerate}
For each $p$, we find $\mathcal{G}(\rho_0(p))$ as well as the GQD of the states which have the greatest fidelities with $\rho_1$ and $\rho_2$. That is to say, we calculate $\max_{\theta_{1,2},\phi_{1,2}}[\mathcal{F}(\rho_{M_1M_2}^\prime(p),\rho_i)]$ for $i=1,2$ and where $\mathcal{F}(\sigma_k,\sigma_l)$ denotes the fidelity between two quantum states $\sigma_k$ and $\sigma_l$. Once the optimal measurement parameters are found by this maximization, we find the GQD of the corresponding state for each $p$. We plot our findings in Fig.~\ref{fig: appendix plots 1}~\textbf{(b)}. If we look at the maximum value of the three plots for each value of $p$, the graph ends up looking remarkably similar to Fig.~\ref{fig: appendix plots 1}~\textbf{(a)}, and so we may attribute the behavior to changing fidelities with Bell state mixtures for different values of $p$.

We also note that the optimal 'target state' (in terms of GQD) for $p=1$ is $\rho_2$, with a fidelity extremely close to 1 with this state when the measurement parameters $\theta_1=\theta_2\approx0.9553$ and $\phi_1=\phi_2=3\pi/4$ are used. However, one change in this modified protocol compared to the original is that the GQD is no longer measurement-outcome independent. For $p=1$ and the measurement basis defined using the above angles, $\rho_2$ is only obtained with a fidelity close to unity when the measurement outcomes are identical, which occurs with probability $\frac{1}{2}$. If orthogonal measurement outcomes are observed, then the memories are left in a state that has a GQD of $\sim0.1258$ corresponding to a very high fidelity with the Werner state $y\ketbra{\Psi^-}{\Psi^-} + \frac{1-y}{4}\Id$ for $y=\frac{1}{3}$. We point out that this state is entangled for $y>\frac{1}{3}$.
% \begin{figure}[t]
%     \centering
%     \includegraphics[width=\linewidth]{x_rho3.pdf}
%     \caption{Optimal $x$ values plotted against $p$. Each optimal $x$ defines the state $\tau(x)$ with which a high fidelity results in the most GQD for each $p$-value.}
%     \label{fig: optimal x}
% \end{figure}

Although this gives an idea as to why Fig.~\ref{fig: appendix plots 1}~\textbf{(a)} exhibits such peculiar behavior, we wish recreate it exactly using the logic used to generate Fig.~\ref{fig: appendix plots 1}~\textbf{(b)}. To do this, we assume that the 'target state' which with a high fidelity will gives the maximum amount of GQD is of the form $\tau(x)$, but with a different $x$ for each $p$. That is to say, each value of $p$ has a different optimal value of $x$ which defines the target state $\tau(x)$; this optimal value of $x$ results in the maximum amount of GQD when the fidelity between the memory state and $\rho_3(x)$ is optimized over carrier measurements. It is not helpful to simply find the value of $x$ which results in the maximum fidelity, as the maximum fidelity doesn't necessarily also result in the maximum GQD.

Employing an optimization algorithm incorporating a grid-search, we obtain the blue plot in Fig.~\ref{fig: appendix plots 2}~\textbf{(a)}. This plot is almost identical to Fig.~\ref{fig: appendix plots 1}~\textbf{(a)} apart from $0\leq p \lessapprox0.09$ in which there are smaller optimal values of GQD. In order to exactly recreate the results, we include the plot of GQD in the $\rho_0(p)$ states from Fig.~\ref{fig: appendix plots 1}~\textbf{(b)}; taking the maximum between the two plots for each $p$ results in a perfect copy of the original graph. This suggests that $p\approx0.09$ is some 'threshold value' of correlated noise, beyond which fidelity with mixtures of Bell states out-performs the structure of states obtained from the original protocol.
We visualize the sudden rise at $p\approx0.12$ further by plotting the optimal value of $x$ against $p$ in Fig.~\ref{fig: appendix plots 2}~\textbf{(b)}. One can see that the optimal 'target state' suddenly changes at this value of $p$, indicating that beyond this value, states close to $\tau(x)$ can both outperform $\rho_0(p)$ and keep increasing with the noise strength.
\end{document}